\documentstyle[12pt]{article}
\input epsf
\def\be{\begin{equation}}
\def\ee{\end{equation}}
\def\l{\label}
\def\refe#1{(\ref{#1})}

\def\ie{{\it i.e.}}

\def\LEP2{{LEPII}}
\def\mg{$m_{3/2}$}
\def\th{$\theta$}

\def\npb#1#2#3{    {\it Nucl. Phys. }{\bf B #1} (19#2) #3}

\def\plb#1#2#3{    {\it Phys. Lett. }{\bf B #1} (19#2) #3}
\def\prd#1#2#3{    {\it Phys. Rev. }{\bf D #1} (19#2) #3}
\def\prep#1#2#3{   {\it Phys. Rep. }{\bf #1} (19#2) #3}
\def\prl#1#2#3{    {\it Phys. Rev. Lett. }{\bf #1} (19#2) #3}

\def\ibid#1#2#3{   {\it ibid. }{\bf #1} (19#2) #3}

\begin{document}
\begin{titlepage}
\vspace*{-1.5cm}
\begin{center}
\vspace{5ex}
{\Large \bf Sparticle And Higgs Masses Within}\\
{\Large \bf Minimal String Unification}\\
\vspace{3ex}
{\bf Shaaban Khalil}$^{a,b}$, {\bf Antonio Masiero}$^{c}$ {\bf and Qaisar 
Shafi}$^{d}$\\
{\it
\vspace{1ex} a) International Center For Theoretical Physics, ICTP,
Trieste, Italy.}\\
{\it
\vspace{1ex} b) Ain Shams University, Faculty of Science, Department of
Mathematics\\
 Cairo, Egypt.}\\
{\it
\vspace{1ex} c) Dipartimento di Fisica, Universit\`{a} di Perugia,and\\
INFN, Sezione di Perugia, Perugia, Italy}\\ 
{\it
\vspace{1ex} d)Bartol Research Institute, University of Delaware\\ 
Newark, DE 19716, USA}\\

\vspace{6ex}
{ABSTRACT}
\end{center}
\begin{quotation}
We consider the sparticle and higgs spectroscopy in a class of 
superstring inspired models in which the string 
threshold corrections ensure the consistency of the string 
unification scale with the low energy data. The lightest neutralino is 
almost a pure bino and it is predicted to be the lightest sparticle 
(LSP). Requiring that $\Omega_{LSP}\leq 0.9$, we find an upper bound on its 
mass which, in the case of dilaton supersymmetry breaking, turns out to be 160 
GeV. The LEP 1.5 experimental lower bound on the chargino mass, 
$m_{\chi^{\pm}} > 65$ GeV, 
implies that the lower bound on the LSP mass is $m_{LSP} > 32 (45)$ GeV, 
corresponding to $\mu < (>)0$. We also determine the lower and upper 
bounds on the higgs and other sparticle masses. For instance, the 
lightest higgs lies between 65 and 115 GeV, while the mass of the 
lightest  charged sparticle satisfies 47 GeV $ < m_{\tilde{e}_R} < 325$ 
GeV. With only the top Yukawa coupling of order unity we find that $ 1.5 \leq 
\tan \beta \leq 3.5 $. \end{quotation} \end{titlepage} 
\vfill\eject

\hspace{0.3cm}By incorporating gravity from the outset, superstring theories 
appear to provide a `more complete' unification of the fundamental forces than 
standard supersymmetric grand unification. It is natural to suspect that the 
string unification scale $M_c$ is comparable to the (reduced) Planck Mass 
$M\sim 2.4\times 10^{18}$ GeV. More concrete calculations suggest that 
$M_c$ is about an order of magnitude below M. This still poses difficulties for the 
theory since it would seem to be inconsistent with the scale $M_{GUT} \sim 
2\times 10^{16}$ GeV deduced from considerations of the low energy data.\\

   A number of ideas have been put forward to tackle this problem,
 including
the existence of a grand unified group, the presence of new particles at
 an intermediate scale or, more recently, the possibility that $M_c$ can indeed
be lowered to coincide with $M_{GUT}$.
One approach for resolving this discrepancy has to do with 
the potential string threshold corrections. This can be realized in certain 
orbifold compactification schemes in which the consistency of the string unification 
scale $M_c$ with the low energy data is achieved by an appropriate 
choice of the modular weights of the matter fields \cite{Ross}. Following
 these authors, we refer to them as "minimal string unification". 
 Moreover, in a special 
class of such models \cite{ibanez1} on which we will focus here, the 
supersymmetry breaking 
soft terms are characterized in terms of the gravitino mass \mg, the modular 
weights $n_i$ and a mixing angle \th\ defined by $ \tan \theta= \langle 
F^S\rangle /\langle F^T \rangle$. Here $\langle F^S \rangle$ and $\langle 
F^T \rangle$ denote the  magnitude of the vacuum expectation values
 (vevs) of the F-terms of 
the dilaton and modulus supermultiplets $S$ and $T$. For simplicity we will 
assume the presence of a single dominant modulus $T$.\\

	An attractive feature of this approach, which was recently emphasized, 
is that it leads to a number of `low energy' predictions which can be tested 
at \LEP2, Tevatron and, of course, the LHC \cite{khalil}. 
In particular, within the framework 
of radiative electroweak breaking it was shown that in the small 
$\tan\beta$ region, the lightest charged sparticles include a chargino 
and the (right) sleptons, and that interesting correlations occur between 
their masses and with the mass of the lightest higgs~\cite{khalil}. 
These should provide important tests of the particular superstring 
inspired scheme.\\

	In this paper we wish to pursue this investigation further by 
focussing in particular on the composition and mass of the LSP. Remarkably, 
the lightest neutralino turns out to be the LSP ( without imposing 
additional requirements), and it is very nearly a pure bino for much of 
the parameter 
range. It annihilates predominantly into lepton - antilepton pairs via 
the `right' slepton exchange. These considerations, and in particular 
the requirement that 
$\Omega_{LSP}\leq 0.9$ ( with $\Omega_{total}=1$ from inflation) lead us 
to a fairly stringent upper bound on the LSP mass which depends on the 
"mixing angle" \th. For  $\theta=\pi/2$ \ie in the pure dilaton case, for 
instance, the upper bound is 
160 GeV, with the maximum value of 300 GeV occuring for \th\ close to 
0.98 radian. Furthermore, this enables us to provide upper bounds on the 
sparticle masses, as well as a lower bound (of $\sim 65$ GeV) on the mass 
of the lightest ($CP$ even) neutral higgs. Note that the lower bound of 
65 GeV by LEP 1.5 on the lightest chargino mass \footnote{This bound is 
increasing with the higher available energy at \LEP2. Presently it has 
reached approximately 84 GeV.} implies a lower bound on the 
LSP mass of 32 GeV which is rather insensitive to the \th\ parameter. Our 
analysis in the context of minimal string unification agrees with some 
previous results obtained by De Carlos and Kraniotis ~\cite{kraniotis} in 
effective supergravities.\\
 
	In ref~\cite{khalil}, following minimal string unification,
 we studied the  phenomenological 
implications of effective supergravity theories derived from string vacua with 
N=1 supersymmetry spontaneously broken by the dilaton and moduli 
$F$-terms~\cite{ibanez1}.  
 The 
mass spectrum in this approach is determined in terms of 
two independent parameters, the 
dilaton-modulus mixing angle \th\ and the 
gravitino mass \mg. The various soft 
supersymmetry breaking terms at the compactification scale are characterized as 
follows. 
The scalar masses are   
\begin{equation}
m^2_i = m^2_{3/2}(1 + n_i \cos^2\theta),
\l{scalar}
\end{equation}
where $n_i$ are the various modular weights given in ref~\cite{Ross} to 
obtain minimal string unification:
$$n_{Q_L}=n_{D_R}=-1,\hspace{0.5cm} n_{u_R}=-2,\hspace{0.5cm}
n_{L_L}=n_{E_R}=-3,\hspace{0.5cm} n_{H_1}=-2,\hspace{0.5cm} n_{H_2}=-3$$
The above modular weights are taken to be family-independent. The asymptotic 
gaugino masses are
\begin{eqnarray}
 M_1 = \sqrt{3} m_{3/2} \left( \sin\theta - 0.02 \cos\theta \right)
\l{m1}\\
 M_2 = \sqrt{3} m_{3/2} \left( \sin\theta + 0.06 \cos\theta \right)
\l{m2}\\
 M_3 = \sqrt{3} m_{3/2} \left( \sin\theta + 0.12 \cos\theta \right)
\l{m3}
\end{eqnarray}
In the soft sector of the trilinear scalar couplings we focus only on the
A-term which is related to the top quark Yukawa coupling $A_t$.
It is given by:
\begin{equation}
A_t = - m_{3/2}(\sqrt{3}  \sin\theta - 3 \cos\theta)
\l{at}
\end{equation}
The bilinear soft breaking
term $B \mu H_1 H_2$ (where $H_1$ and $H_2$ denote the scalar doublets)
depends on the origin of the $\mu$-term in the
superpotential. If $\mu$ arises solely from the $S$ and $T$ 
sector then, as pointed out in ref~\cite{ibanez1}, $B$ takes the form:
\begin{equation}
B = m_{3/2}(-1 - \sqrt{3}  \sin\theta + 2 \cos\theta)
\l{b}
\end{equation}

	Given the boundary conditions in equations \refe{scalar} to \refe{b} at the
compactification scale, we determine the
evolution of the `couplings' according to their one loop renormalization group  
equations (RGE) in order to estimate the mass spectrum of the supersymmetric 
particles at the weak scale.
	The radiative electroweak symmetry breaking scenario imposes the 
following well known conditions among the
renormalized quantities: \\
\be
m_{H_1}^2 +m_{H_2}^2+2 \mu^2 > 2B \mu, 
\ee
\be
(m_{H_1}^2+\mu^2)(m_{H_2}^2+\mu^2)<(B\mu)^2
\ee
and
\be
\mu^2 = \frac{ m_{H_1}^2 -m_{H_2}^2 \tan^2\beta}{\tan^2\beta - 1}
- \frac{M_Z^2}{2},
\ee
\be
\sin 2\beta= \frac{-2  B \mu }{m_{H_1}^2+m_{H_2}^2+ 2\mu^2 },
\l{minimization}
\ee
where $\tan\beta= {\langle H_2^0 \rangle}/{\langle H_1^0 \rangle}$ is the 
ratio of the two higgs vevs that give masses to the up and down type 
quarks and $m_{H_1}$, $m_{H_2}$ are the two higgs masses at the 
electroweak scale.  

	Using equations \refe{scalar}-\refe{minimization} we find that $\mu$ and 
$\tan \beta $ are specified in terms
of the goldstino angle $\theta$ and the gravitino mass \mg. As is the 
case in MSSM, in our scheme too the results 
are in general sensitive to the sign of $\mu$. We find that the choice of 
negative $\mu$ allows for a lighter sparticle spectrum, in particular 
 the chargino and the right selectron. In view of 
our special interest in potential discoveries at \LEP2, we adopt this choice for the 
sign of $\mu$ in the present analysis. Also, for $\tan \beta$ we prefer 
 to work in the low $\tan \beta$ regime, \ie in the RGE evolution we 
consider that all Yukawa couplings, except $h_t$, are much smaller than unity.
Interestingly enough, it turns out that  
 only a rather narrow interval is allowed, namely 
$1.5 \leq \tan \beta \leq 3.5$. As explained in ~\cite{khalil}, an 
important  constraint on $\theta$
 arises from the requirement  of
colour and especially electric charge conservations\footnote{We impose 
the coservative constraint of the absence of electric charge or colour 
breaking local minima. Hence we do not take into account the possibility 
of existence of other nearby minima which conserve elecric charge and 
colour and are reachable by tunnelling in a cosmologically short enough 
time.}, 
namely $ 0.98 rad. < \theta < 2 rad$ (the upper bound is not sensitive to
the sign of $\mu$ ). \\

	The neutralinos $\chi_i^0 $ (i=1,2,3,4) are
the physical (mass) superpositions of the Higgsinos $\tilde{H}_1^0$,  
$\tilde{H}_2^0 $, and the two neutral gauginos 
$\tilde{B}^0$ (bino) and $\tilde{W}^0_3$ (wino). The neutralino mass 
matrix is given by~\cite{nilles} 
\be M_N = \left(\begin{array}{clcr}M_1&0 &
-M_Z\cos\beta\sin\theta_W &M_Z\sin\beta\sin\theta_W\\
 0&M_2&M_Z\cos\beta\cos\theta_W&-M_Z\sin\beta\cos\theta_W\\
-M_Z\cos\beta\sin\theta_W &   
M_Z\cos\beta\cos\theta_W&0&\mu\\
 M_Z\sin\beta\sin\theta_W&-M_Z\sin\beta\cos\theta_W&\mu&0
\end{array}\right)
\l{neutralino}
\ee
where $M_1$ and $M_2$ now refer to `low energy' quantities whose asymptotic 
values are given in equations \refe{m1} and 
\refe{m2}. The lightest eigenstates $\tilde{\chi}^0_1$ is a linear 
combination of the original fields:
\be
\tilde{\chi}^0_1 = N_{11}\tilde{B}+ N_{12}\tilde{W}^3+
N_{13}\tilde{H}_1^0 + N_{14}\tilde{H}_2^0,
\ee
where the unitary matrix $N_{ij}$ relates the $\tilde{\chi}^0_i$ fields to
the original ones. The entries of this matrix depend on
$\tan \beta$, $M_2$ and $\mu$ which,  as previously mentioned, are 
determined in terms of \th\ and \mg. The dependence of 
the $\tilde{\chi}^0_1$ (LSP) mass on \mg\ is 
shown in figure 1. A useful parameter for describing the neutralino 
composition is the gaugino "purity" function~\cite{report} 
\be
f_g= \vert N_{11}\vert^2 + \vert N_{12} \vert^2
\ee
We plot this function versus \mg\ in figure (2) which clearly shows 
that the LSP is essentially a pure bino.\\

	Given the LSP mass as a function of \mg\ and \th\ 
and that it is bino like, we find that the annihilation is predominantly 
into leptons, with the other channels~\cite{report} either  
closed or suppressed. The annihilation process is dominated 
by the exchange of the three slepton families 
($\tilde{e}_R$,$\tilde{e}_L$, etc). The squark exchanges 
are suppressed due to their large mass, while the Z-boson 
contribution is suppressed, except for $m_{\chi} \sim m_Z/2$, due to the 
small $Z \chi \chi$ coupling
$ (ig/2 \cos\theta_W) (N_{13}^2-N_{14}^2) \gamma^{\mu} \gamma_5$.

	For the computation of the lightest neutralino relic abundance  we
follow the standard procedure~\cite{ellis}. First we need to determine the
thermally averaged cross section $\langle \sigma_A v 
\rangle \sim a + b v$~\cite{grist}. Since the lepton masses are small 
compared to $m_{LSP}$, we find that $a\sim 0$, while b is given by:
\begin{eqnarray}
b &=& \sum_{\tilde{l_R},\tilde{l_L}} \frac{4}{\pi} G_F^2 
m_{\chi}^2 y'^4 (u'^2+v'^2)(\frac{2}{3}+r_1) + 
\frac{1}{4}(N_{13}^2-N_{14}^2)^2 x'^4 (c_L^2 + c_R^2)
\nonumber\\
&+&\frac{2}{3} (N_{13}^2-N_{14}^2) x'^2 y'^2 \left((v' c_R -u' c_L) -
r (u' c_L +v' c_R) \right)
\end{eqnarray}
where $G_F$ is the Fermi constant, $r_1= \frac{r}{3} 
(-4 + 4 r )$, with
$$r=\frac{m_{\chi}^2}{M_{\tilde{l}}^2+m_{\chi}^2},$$  
$$ y'^2 = \frac{m_W^2}{M_{\tilde{l}}^2+m_{\chi}^2},$$ 
and
$$x'^2=\frac{m_Z^2}{((m_Z^2-s)^2+\Gamma_Z^2 m_Z^2)^{1/2}}$$ is the $Z$ pole 
factor with $\Gamma_Z$ the $Z$ decay width. Finally, 
$$u'= (T_{3L}N_{12}-\tan \theta_W(T_{3L}-e_l)N_{11})^2 , \hspace{1.5cm}  v'= 
(\tan \theta_W e_l N_{11})^2.$$
Here $T_{3L}$ is the weak isospin, $e_l$ is 
the lepton charge, $\sin^2\theta_W=0.23$, 
$c_L=T_{3L}-e_l \sin^2 \theta_W$ and $c_R=-e_l \sin^2\theta_W$. \\

	Given $a$ and $b$ we can determine the freeze-out temperature $T_F$, 
below which the $\chi \chi $ annihilation rate is smaller than the expansion rate of 
the universe. Following refs(~\cite{ellis},~\cite{kolb} and ~\cite 
{grist}) we can iteratively compute the freeze-out temperature from
\be
x_F= \ln \frac{0.0764 M_P ( a+ 6 b/ x_F) c ( 2+c) m_{\chi}}{\sqrt{g_* x_F}}
\ee
Here $ x_F=m_{\chi}/T_F$, $ M_P= 1.22 \times 10^{19} $ GeV is the Planck 
mass and $ g_*$ ($ 8 \leq \sqrt{g_*} \leq 10$) is the effective number of 
relativistic degrees of fredom at $T_F$. Also $c=1/2$ as explained in 
Ref~\cite{nojiri}.

	The relic LSP density is given by
\be
\Omega_{\chi} h^2 = \frac{\rho_{\chi}}{\rho_c/h^2} = 2.82 \times 10^8 
Y_{\infty} (m_{\chi}/GeV),
\ee
where
\be
Y_{\infty}^{-1}=0.264\ g_*^{1/2}\ M_P\ 
m_{\chi}\ (\frac{a}{x_F}+\frac{3b}{x_F^2}),
\ee 
$h$ is the well known Hubble parameter, $ 0.4 \leq h \leq 0.8$, and $ 
\rho_c \sim 2 \times 10^{-29} h^2$ is the critical density of the 
universe. Motivated by the inflationary scenario we will assume that the 
total density parameter $\Omega_{TOT}=1$. \\

	Figure 3 shows the relic abundance of the lightest neutralino 
$\Omega_{\chi}h^2$ as function of the gravitino mass in the case of  
dilaton supersymmetry breaking, \ie\ $ \theta=\pi/2$. We require the neutralino relic 
density to be $ 0.1 \leq \Omega_{LSP} \leq 0.9$, with $ 0.4\leq h 
\leq 0.8$. We find that there is no point in the parameter space (\mg\ 
, \th\ ) that leads to $\Omega_{\chi} h^2$ less than the minimum value 
(0.014), while the maximum value (0.576) imposes an upper bound on \mg\ 
which is very sensitive to \th\ , as shown in figure 4. In turn, 
this leads to ( a fairly stringent) upper 
bound on the LSP mass of about 160 GeV. 
 For $\theta \simeq 0.98 rad. $, which 
represents the maximum moduli contibution to SUSY 
breaking, \ie\ maximal non-universal soft SUSY breaking terms, this  
bound approaches 300 GeV.\\

	The bounds on the gravitino mass can be translated into bounds on 
the sparticle masses. We are particularly interested in the particles 
which possibly can be seen at \LEP2, Tevatron or the LHC. These include 
the lightest higgs scalar ($h^0$) as well as the "right" selectron and the 
lightest chargino.
	At tree level, the mass of  $h^0$ is 
determined by 
\be
m^2_{h^0}= \frac{1}{2} \left( m_A^2+m_Z^2 - \sqrt{(m_A^2+m_Z^2)^2- 4 m_Z^2 
m_A^2 \cos^2 2\beta} \right)
\ee
where $m_A^2=m^2_{H_1}+m^2_{H_2}+2 \mu^2$.
	The leading  radiative corrections to $m_{h^0}^2$ depend on the fourth 
power of the top mass as well as on  
 $m_{\tilde{t}_{1,2}}$, $A_t$, $\mu$ and $\tan\beta$. The expression for the 
lightest higgs mass which we used in our calculation has the form ~\cite{haber}
\be
m_h^2=m_{h^0}^2 + (\Delta m_h^2)_{1LL} + (\Delta m_h^2)_{mix}
\l{1lop}
\ee
where
\be
(\Delta m_h^2)_{1LL}= \frac{3 m_t^4}{4\pi^2 v^2} 
\ln(\frac{m_{\tilde{t}_1}m_{\tilde{t}_2}}{m_t^2}) 
\left[1+O(\frac{m_W^2}{m_t^2})\right] \ee
\be
(\Delta m_h^2)_{mix}= \frac{3 m_t^4 \tilde{A}_t^2}{8\pi^2 v^2}
\left[ 2h(m_{\tilde{t}_1}^2,m_{\tilde{t}_2}^2)+ \tilde{A}_t^2 
f(m_{\tilde{t}_1}^2,m_{\tilde{t}_2}^2)\right] \left[1+O(\frac{m_W^2}{m_t^2})\right] 
\ee
and $\tilde{A}_t=A_t +\mu \cot \beta$\ \footnote{We use the sign convention 
of $\mu$ opposite to that adopted in the Haber 
and Kane report~\cite{nilles}}, where 
the functions $h$ and $f$ are given by \be
h(a,b) = \frac{1}{a-b} \ln(\frac{a}{b}) \hspace{1cm} and \hspace{0.2cm} 
f(a,b)=\frac{1}{(a-b)^2}\left[2-\frac{a+b}{a-b} \ln(\frac{a}{b})\right]
\ee

	This expression provides the upper bound on $m_h$ for a given 
stop spectrum which is completely determined in terms of \mg\ and \th. 
However including two-loop effects remains necessary to obtain a correct 
estimate of the higgs mass. It was shown in Ref~\cite{haber} that the two 
loop leading logarithmic contributions to $m_h^2$ are incorporated by 
replacing $m_t$ in equation \refe{1lop} by the running top quark mass 
evaluated at the scale $\mu_t$ which is given by $\mu_t =\sqrt{M_t M_s}$ 
where $M_t$ is the pole mass of the top quark and 
$M_s=\sqrt{\frac{M_{\tilde{t_1}}^2+M_{\tilde{t_2}}^2}{2}}$.
Then the lightest higgs mass is given by
\be
m_h^2=m_{h^0}^2 + (\Delta m_h^2)_{1LL}(m_t(\mu_t)) + (\Delta 
m_h^2)_{mix}(m_t(M_s)) 
\l{2lop}
\ee     

	Figure 5 shows how the higgs mass varies with \mg\ and \th, and we see 
that in the low $\tan \beta $ regime and with $\mu > 0$ ( which gives 
maximum mixing in our convention),  the  mass $m_h$ satisfies $80 GeV \leq 
m_h \leq 115 GeV$. The lower bound reaches 65 Gev in the case of $\mu 
<0$ and $\theta=2$ rad. \\
 
We can similarly determine both the lower and upper bounds on all the 
supersymmetric particles since they are given in terms of the goldstino 
angle \th\ and the gravitino mass \mg. For instance, we find that the 
mass of the pseudoscalar A lies between 200 GeV and 2 TeV, while the gluino
mass is between 400 GeV and 2.5 TeV.\\

	In conclusion, we have discussed how the composition and cosmic 
abandance of the LSP in the so-called minimal superstring unification leads to 
important constraints on the underlying supersymmetry breaking parameters.
In addition, one finds lower and upper bounds on the higgs and sparticle
mass spectrum. For instance, the "Weinberg-Salam" higgs is estimated to lie 
in the mass range of 85-145 GeV. The squarks and gluinos turn out to be 
heavy but, depending on the parameters, a charged slepton or a chargino 
could still be found at \LEP2.
\vskip0.1truecm

\noindent{\Large\bf Acknowledgements} 
\vskip0.5truecm
 The authors would like to acknowledge the hospitality of ICTP where this
work was initiated. One of us (Q.S) acknowledges support by the US Departmet
of Energy, Grant No. DE-FG02-91ER 40626.

\newpage
\noindent{\Large\bf Figure Captions}
\vskip0.5truecm
\noindent{\bf Fig.\ 1} The lightest neutralino mass as function of the 
gravitino mass (\mg\ ). \\
{\bf Fig.\ 2} The gaugino `purity' function versus the gravitino mass.\\
{\bf Fig.\ 3} The neutralino relic abundance as a function of the
gravitino mass in the pure dilaton case of supersymmetry breaking. The 
long- and short- dashed lines correspond to 
$\Omega_{LSP}=0.1$, $h=0.4$ and $\Omega_{LSP}=0.4$, $h=0.8$, respectively\\
{\bf Fig.\ 4} The allowd region of the parameter space (\mg\ , \th\ )
corresponding to $0.1\leq \Omega_{LSP}\leq 0.9$ and $m_{\chi_{\pm}} 
\geq 65 GeV (\mu < 0 ) $.\\
{\bf Fig.\ 5} The lightest (`Weinberg-Salam') higgs mass as function of the
 gravitino mass.\\
\vfill
\eject
\begin{figure}
\epsfxsize=\hsize
\epsffile{ 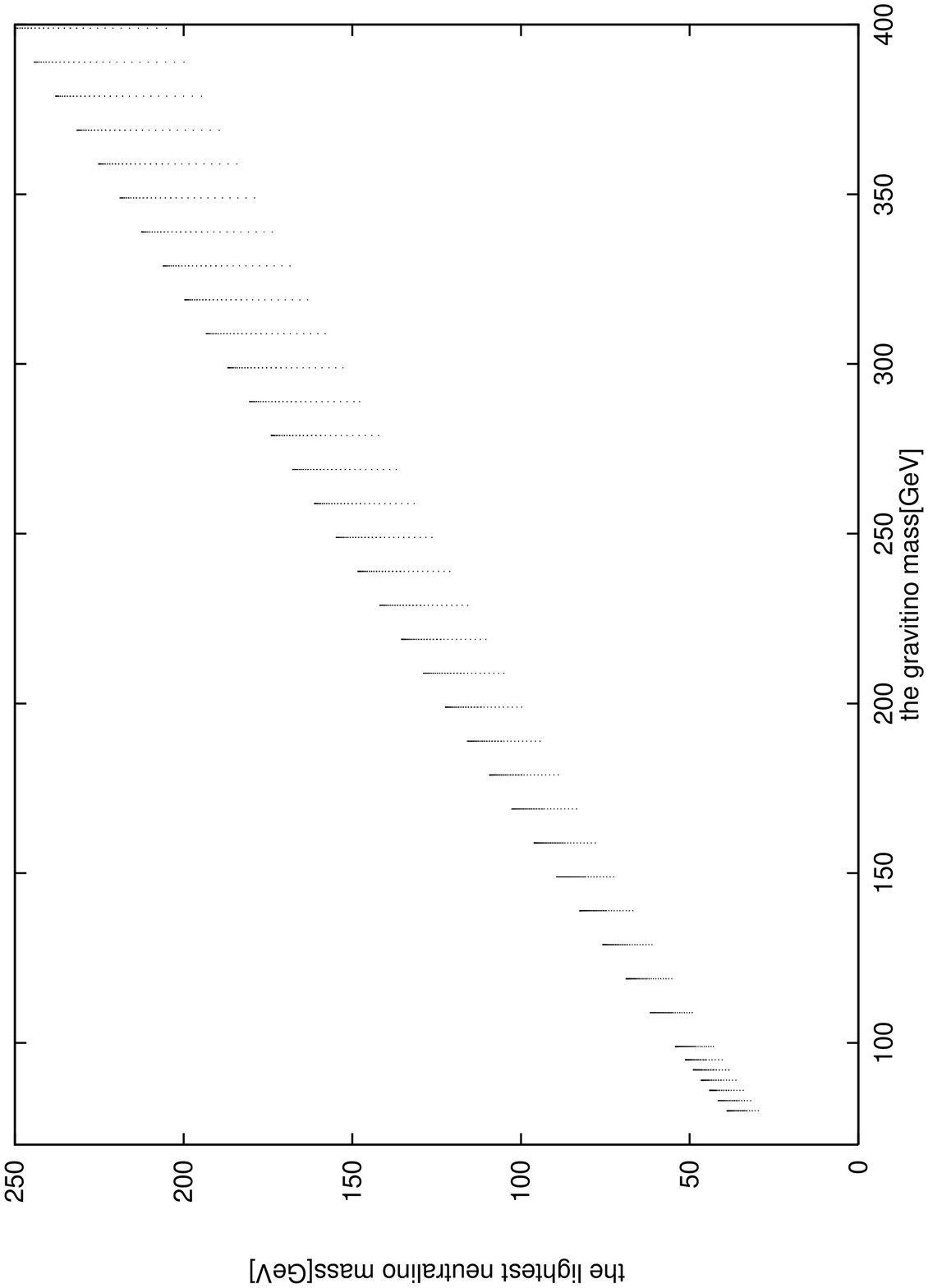}
\end{figure}
\begin{figure}
\epsfxsize=\hsize
\epsffile{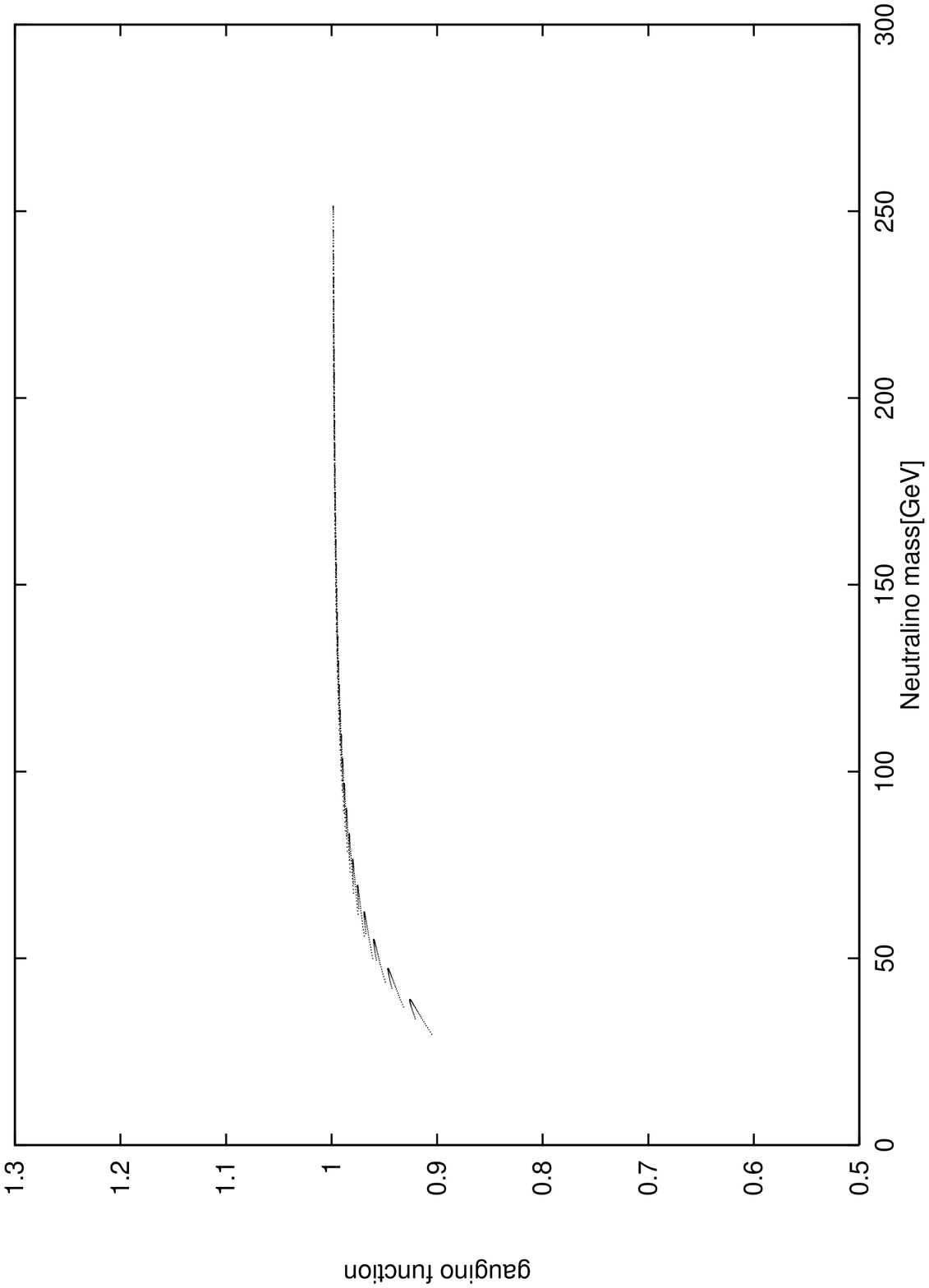}
\end{figure}
\begin{figure}
\epsfxsize=\hsize
\epsffile{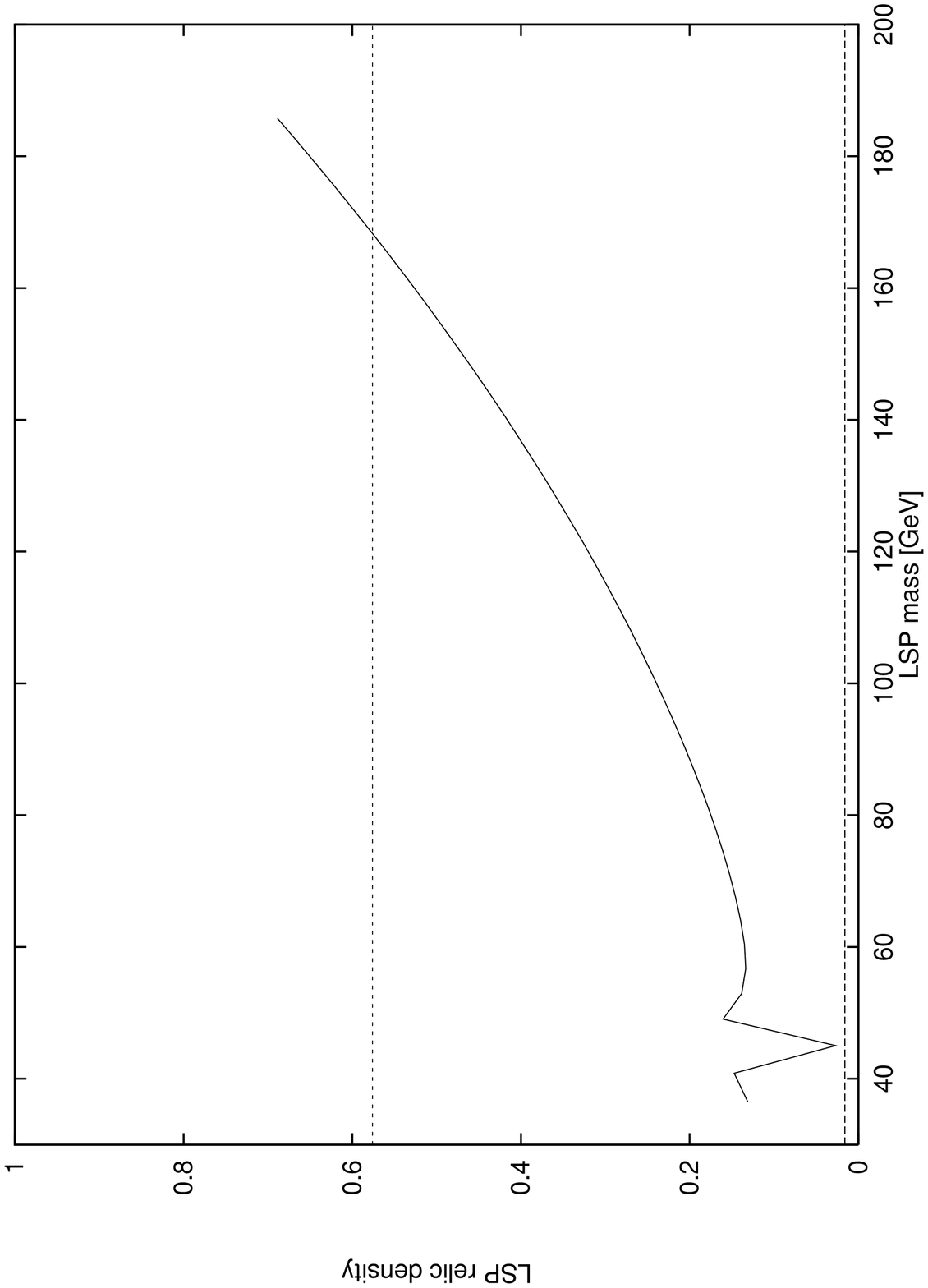}
\end{figure}
\begin{figure}
\epsfxsize=\hsize
\epsffile{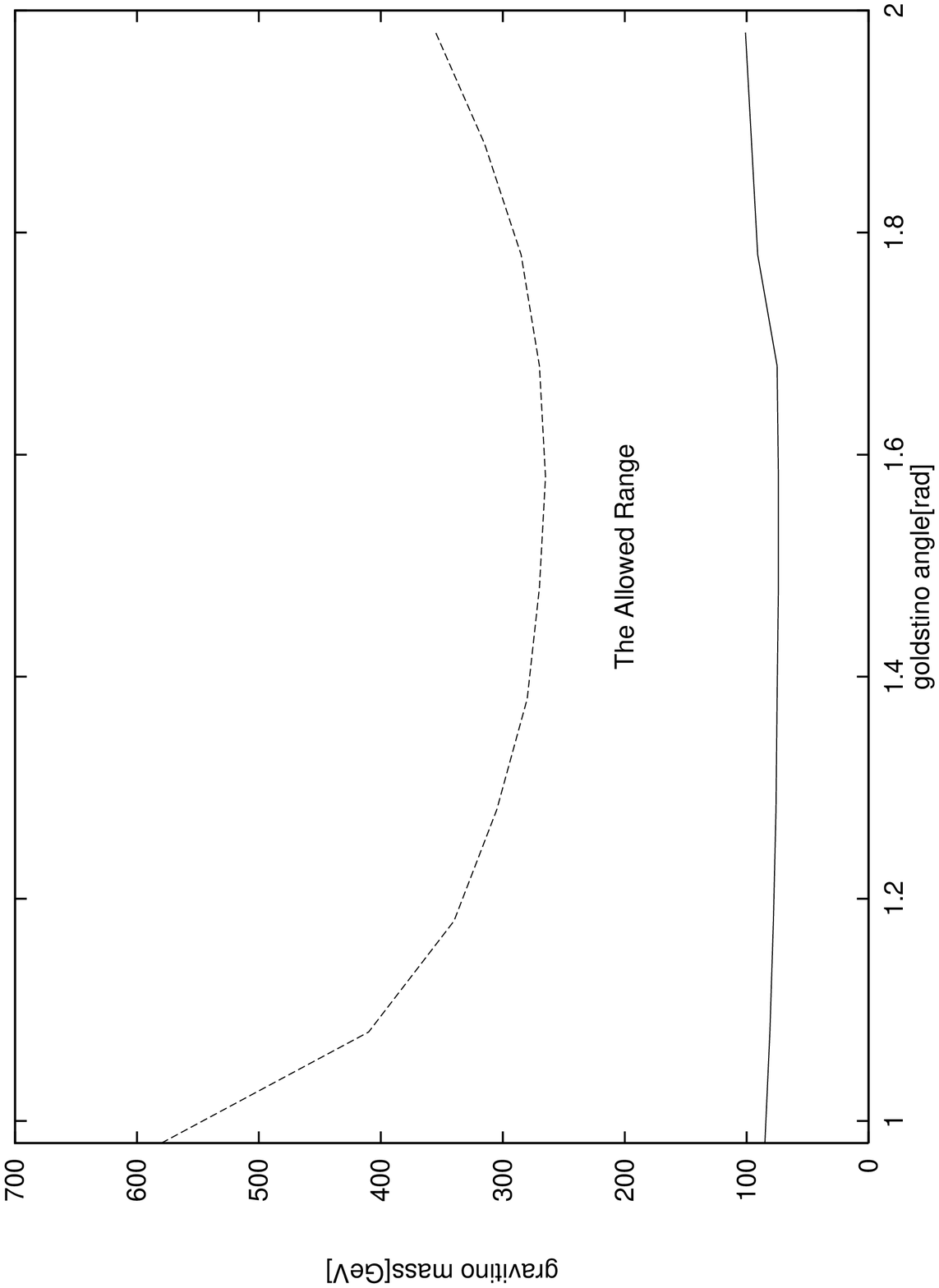}
\end{figure}
\begin{figure}
\epsfxsize=\hsize
\epsffile{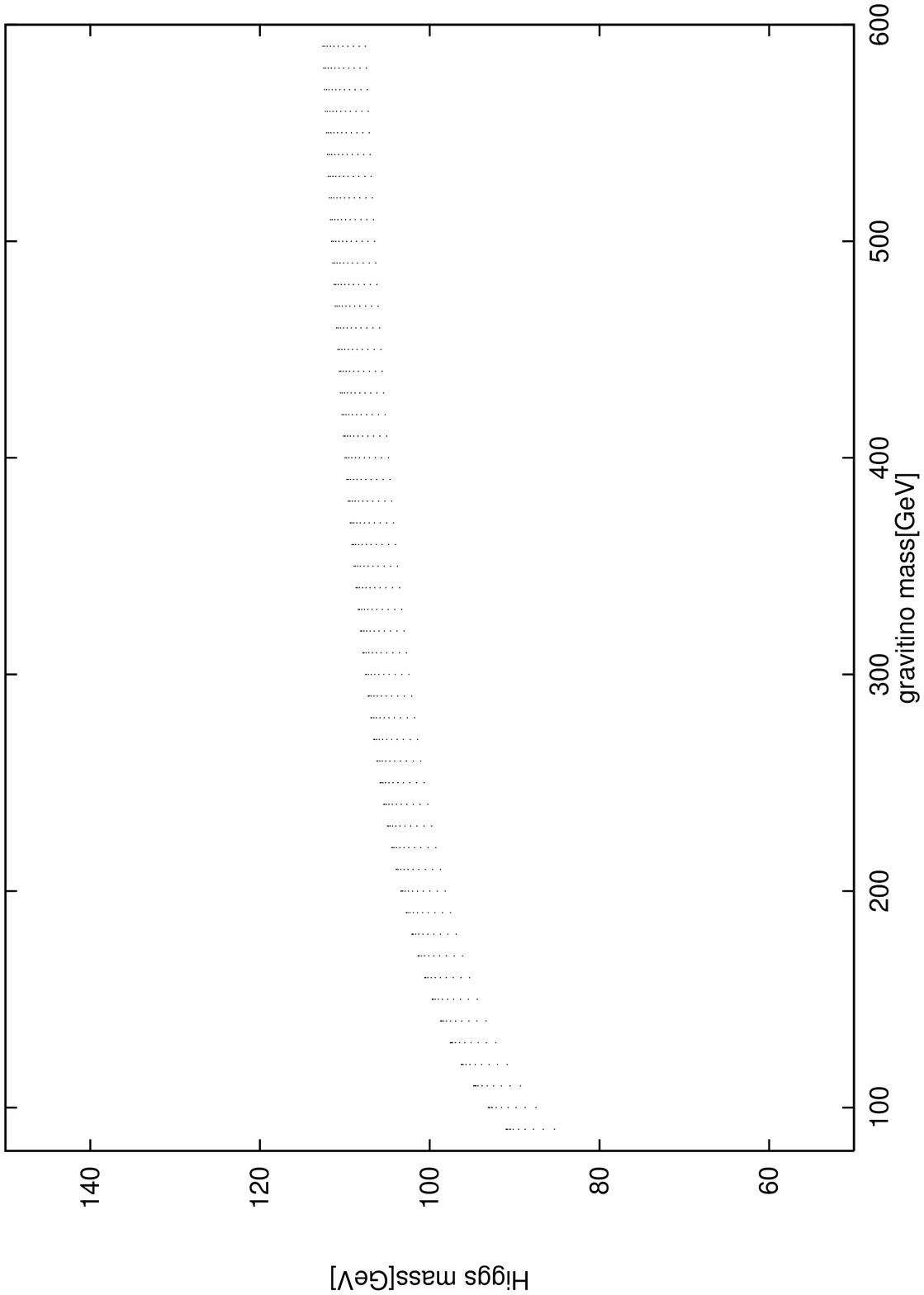}
\end{figure}
\end{document}